\documentclass[11pt,onecolumn,amssymb,nofootinbib]{revtex4}
\usepackage{amsmath, amsthm, amscd, amssymb}
\usepackage{bm}
\usepackage{bbm}
\usepackage{slashed}
\usepackage{graphicx}
\usepackage{hyperref}

\begin{document}

\title{\bf Chiral Anomaly Calculation in the Extended Coupled Rarita-Schwinger Model}

\author{Stephen L. Adler}
\email{adler@ias.edu} \affiliation{Institute for Advanced Study,
Einstein Drive, Princeton, NJ 08540, USA.}

\author{Pablo Pais}
\email{pais@ipnp.troja.mff.cuni.cz} \affiliation{Faculty of Mathematics and Physics, Charles University, V Hole\v{s}ovi\v{c}k\'ach 2, 18000 Prague 8, Czech Republic
}

\begin{abstract}
We recalculate the chiral anomaly in the Abelian gauge model in which a spin-$\frac{1}{2}$ field is directly coupled to a Rarita-Schwinger spin-$\frac{3}{2}$ field, using the extended theory in which there is an exact fermionic gauge invariance. Since the standard gauge fixing and ghost analysis applies to this theory, the ghost contribution to the chiral anomaly is $-1$ times the standard chiral anomaly for  spin-$\frac{1}{2}$. Calculation of the fermion loop Feynman diagrams contributing to the coupled model anomaly gives a result of $6$ times the standard anomaly, so the total anomaly is $5$ times the standard anomaly. This agrees with the result obtained from the unextended model taking the ghost contribution there as $0$, corresponding to a non-propagating ghost arising from exponentiating the second class constraint determinant, together with the fermion loop anomaly contribution in the unextended model of $5$ times the standard anomaly.
\end{abstract}

\maketitle

\section{Introduction}
\label{section_Introduction}

In this paper we continue an ongoing investigation of whether Rarita-Schwinger spin-$\frac{3}{2}$ theory can be consistently gauged when not coupled to supergravity.
For the minimal gauged spin-$\frac{3}{2}$ theory, Adler \cite{adler1} showed that by introducing an auxiliary field the theory can be extended to have an exact
fermionic gauge invariance.   A subsequent study of this model by Adler, Henneaux and Pais \cite{ahp} showed that when gauge-fixed in radiation gauge, the extended theory  has an auxiliary field Dirac bracket that is singular for small gauge fields, ruling out a perturbative analysis.  Motivated in part by this,  Adler \cite{adler2} then
studied a model in which a Rarita-Schwinger field is directly coupled to a spin-$\frac{1}{2}$ fermion field, in which the weak field singularity is removed, and a
perturbation theory calculation of the chiral anomaly is possible.  For this coupled model, the chiral anomaly arising from fermion triangle diagrams was found to
be 5 times the standard spin-$\frac{1}{2}$ anomaly.  Since the constraints in the extended model are second class instead of first class, the familiar Faddeev-Popov (FP)
analysis for gauge-fixed first class constraints does not apply, raising questions as to how the ghost contribution to the anomaly should be calculated.  If one proceeds in analogy with the FP method by introducing a ghost field to exponentiate the second class constraint determinant, one finds a non-propagating ghost with an anomaly contribution of 0.  On the other hand, if one adopts a heuristic limiting procedure which supplies a kinetic term to this ghost, making it a propagating ghost, then an anomaly contribution of $-1$ is obtained.  The paper \cite{adler2} did not attempt a definitive choice as to which  of these two possible answers for the ghost contribution to the coupled model anomaly is the correct one.

The purpose of this paper is to combine ideas of \cite{adler1} and \cite{adler2}, by adding an auxiliary field to the coupled model, giving an {\it extended coupled model}
in which there is an exact fermionic gauge invariance.  The standard Rartita-Schwinger fermionic gauge fixing can then be applied, with the ghost contribution giving
an anomaly contribution of $-1$ times the standard spin-$\frac{1}{2}$ chiral anomaly.  The Feynman rules now include vertices linking the spin-$\frac{3}{2}$ and
spin-$\frac{1}{2}$ fermion fields to the auxiliary field, and  propagators for these fields, giving rise to new Feynman diagrams not encountered in the unextended
coupled model.  We find that the one Feynman diagram in the extended model that is analogous to the anomalous triangle in the unextended model has an anomaly of
$6$ times the standard spin-$\frac{1}{2}$ chiral anomaly, while the new diagrams not encountered before in \cite{adler2} all have zero anomalies, giving a total anomaly
of 5 times the standard spin-$\frac{1}{2}$ chiral anomaly.  This agrees with the answer obtained in \cite{adler2} from a non-propagating ghost, and rules out the alternative answer obtained also in \cite{adler2} from the heuristic limiting procedure corresponding to a propagating ghost.

This paper is organized as follows.  In Sec. \ref{section_covariant_left_right} we give the covariant and left chiral forms of the action in the extended coupled model, and show that
the constraint bracket vanishes, as expected for first class constraints. In Sec. \ref{section_path_integral} we give the path integral for this action,  discuss the Nielsen \cite{n}
gauge fixing and ghost anomaly calculation, expand the Lagrangian density in powers of the gauge coupling $g$, and relate this to the formally conserved Noether
currents.  In Sec. \ref{section_Feynman_rules} we give the Feynman rules for propagators and vertices of the extended coupled model. In Sec. \ref{section_enumeration_diagrams} we enumerate the types of Feynman diagrams contributing to the total three point function for the left chiral current in the model, and sketch the strategy for our evaluation of them. In Sec. \ref{section_anomaly_triangle} we calculate the anomaly associated with the triangle in which three leading order Noether currents appear at the vertices, and
show that it is equal to 6 times the standard spin-$\frac{1}{2}$ chiral anomaly, and that it is exactly independent of both the gauge fixing parameter $\zeta$ and
the coupling mass $m$ appearing in the action.  Thus the calculation can also be done by dropping terms which vanish as $\zeta \to \infty$ and $m \to \infty$,
rather than verifying that these cancel among themselves, and keeping only terms which have a finite remainder in this double limit.  In Sec. \ref{section_remaining_diagrams} we apply this idea to  the new diagrams that appear in the extended coupled model, and show that the only terms which survive in the double limit vanish by the identity given in Eq. (72) of \cite{adler2}, and so the new diagrams add 0 to the total anomaly.  Brief conclusions are given in Sec. \ref{section_Discussion}.  Our metric and gamma matrix conventions, and some useful identities, are given in Appendix \ref{appendix_A}.

\section{Covariant and Left Chiral Actions}
\label{section_covariant_left_right}

The covariant form of the action for the extended coupled model will be used to derive path integrals.  It is
\begin{align}\label{covaction}
S=&\int d^4x {\cal L} =S(\psi,\Lambda)+S(\lambda)+S_{\rm interaction}~~~,\cr
S(\psi,\Lambda)=&i\int d^4x \epsilon^{\mu\eta\nu\rho}[\overline{\psi}_{\mu}\gamma_5\gamma_{\eta}D_{\nu}\psi_{\rho}+(g/2)(-\overline{\Lambda}\gamma_5\gamma_{\eta}F_{\mu\nu} \psi_{\rho}+\overline{\psi}_{\mu}\gamma_5\gamma_{\eta}F_{\nu\rho}\Lambda-\overline{\Lambda}\gamma_5\gamma_{\eta}
F_{\nu\rho}D_{\mu}\Lambda)]~~~\cr
S(\lambda)=&-\int d^4x \overline{\lambda} \gamma^{\nu}D_{\nu}\lambda~~~,\cr
S_{\rm interaction}=&m\int d^4x(\overline{\lambda}\gamma^{\nu}\psi_{\nu}-\overline{\psi}_{\nu}\gamma^{\nu}\lambda+\overline{\lambda}\gamma^{\nu}D_{\nu}\Lambda
-\overline{\Lambda}\overleftarrow{D}_{\nu} \gamma^{\nu}\lambda)~~~,\cr
\end{align}
where $D_{\nu}=\partial_{\nu}+gA_{\nu}$, $\overleftarrow{D}_{\nu}=\overleftarrow{\partial}_{\nu}-gA_{\nu}$, so that $A$ is anti-self-adjoint, and
$F_{\mu\nu}=\partial_\mu A_\nu-\partial_\nu A_\mu$. Here $\psi_{\rho}$ is the Rarita-Schwinger field, $\lambda$ is the spin-$\frac{1}{2}$ field to which
the Rarita-Schwinger field is coupled with coupling mass $m$, and $\Lambda$ is the auxiliary field introduced in \cite{adler1} that restores exact fermionic gauge invariance.

The corresponding left chiral form is convenient for studying the constraint structure; it is
\begin{align}\label{leftchiralaction}
S=&S(\Psi,L)+S(\ell)+S_{\rm interaction}~~~,\cr
S(\Psi,L)=&\int d^4x [-\Psi_0^{\dagger} \vec \sigma \cdot \vec D \times \vec{\Psi}+\vec{\Psi}^{\dagger} \cdot
(\vec \sigma \times \vec{D}\Psi_0+\vec{D}\times \vec{\Psi}-\vec \sigma \times D_0 \vec{\Psi})\cr
-&ig\vec{\Psi}^{\dagger} \cdot \vec C L + ig L^{\dagger}\vec C \cdot \vec{\Psi}
+ig \Psi_0^{\dagger} \vec \sigma \cdot \vec B L-ig L^{\dagger}  \vec \sigma \cdot \vec B \Psi_0
+ig L^{\dagger} \vec C \cdot \vec D L-ig L^{\dagger}\vec \sigma \cdot \vec B D_0 L]~~~,\cr
S(\ell)=&i\int d^4x \ell^{\dagger} (D_0-\vec \sigma \cdot \vec D) \ell~~~,\cr
S_{\rm interaction} = &im \int d^4x [-\ell^{\dagger} \Psi_0+ \ell^{\dagger} \vec \sigma \cdot \vec{\Psi} + \Psi_0^{\dagger} \ell -\vec{\Psi}^{\dagger} \cdot \vec \sigma \ell -\ell^{\dagger} D_0L+\ell^{\dagger} \vec \sigma \cdot \vec D L + L^{\dagger} \overleftarrow{D}_0 \ell - L^{\dagger} \overleftarrow{D}\cdot \vec \sigma \ell]~~~,\cr
\end{align}
where $\Psi_{\mu}=P_{L}\psi_{\mu}$, $\ell=P_{L}\lambda$, and $L=P_{L}\Lambda$, with $P_{L}=\frac{1}{2}(1+\gamma_{5})$ the left chiral projector, and
with $\vec C=\vec B+\vec \sigma \times \vec E$.  Writing $S=\int d^4x(-\Psi_0^{\dagger}\chi-\chi^{\dagger}\Psi_0 +...)$, we identify the
constraints as
\begin{align}\label{constraints}
\chi=&\vec \sigma \cdot \vec D\times \vec{\Psi}-im\ell-ig \vec \sigma \cdot \vec B L~~~,\cr
\chi^{\dagger}=&\vec{\Psi}^{\dagger} \times \vec \sigma \cdot \overleftarrow{D} + im \ell^{\dagger} + ig L^{\dagger} \vec \sigma \cdot \vec B~~~.\cr
\end{align}
From Eq. \eqref{leftchiralaction} we can read off the canonical momenta,
\begin{align}\label{canmom}
\vec P_{\vec \Psi}=&\vec \Psi^{\dagger} \times \vec \sigma ~~~,\cr
P_{\ell}=& -i\ell^{\dagger} +imL^{\dagger}~~~,\cr
P_{L}=& ig L^{\dagger} \vec \sigma \cdot \vec B + im\ell^{\dagger}~~~.
\end{align}
Solving for the adjoint fields in terms of the canonical momenta, we get
\begin{align}\label{inverse}
\vec{\Psi}^{\dagger}=&\frac{1}{2}(i\vec P_{\vec \Psi}-\vec P_{\vec \Psi} \times \vec\sigma)~~~,\cr
L^{\dagger}=&-i(P_L+m P_{\ell}) (m^2 + g\vec \sigma \cdot \vec B)^{-1}~~~,\cr
\ell^{\dagger}=&(iP_{\ell}g \vec \sigma \cdot \vec B-im P_L) (m^2 + g\vec \sigma \cdot \vec B)^{-1}~~~.\cr
\end{align}
Using these, and the standard canonical brackets, we find that the bracket of the constraints vanishes,
\begin{equation}\label{constraintbrac}
[\chi_{\alpha}(\vec x),\chi^{\dagger}_{\beta}(\vec y)]=0~~~,
\end{equation}
showing that as expected, in the extended coupled model the constraints have become first class.  It is easy to show that
$\chi$ and $\chi^{\dagger}$ generate the fermionic gauge invariance of the extended model.

\section{Path Integral}
\label{section_path_integral}

Returning to the covariant form, the Feynman path integral is given by
\begin{align}\label{pathint}
<{\rm out}|S|{\rm in}>= &\int \delta(\phi)\delta(\phi^{\dagger})\big[\det[\phi,\chi^{\dagger}] \det[\phi^{\dagger},\chi]\big]^{-1}  d\psi_{\mu}d\psi_{\mu}^{\dagger} d\lambda d\lambda^{\dagger} d\Lambda d\Lambda^{\dagger}\exp(i\int d^4x {\cal L} )~~~,\cr
\end{align}
where $\phi$ and its adjoint $\phi^{\dagger}$ are the constraints introduced to break the fermionic gauge invariance.  If one takes
$\phi=\Lambda$, the left chiral part of the bracket $[\phi,\chi^{\dagger}]$ is $[L,P_{\vec\Psi}\cdot \overleftarrow{D}+P_L]=-1$ and similarly
for the right chiral part, so the Faddeev-Popov (FP) determinant is trivial and there is no ghost contribution; with this choice of gauge fixing the extended
theory reduces to the unextended one discussed in \cite{adler2}.  We will here be interested in Nielsen's choice $\phi=\gamma^{\rho}\psi_{\rho}-b$, with $b$ Gaussian averaged \cite{n}; in this case the factors $\delta(\phi)\delta(\phi^{\dagger})\big[\det[\phi,\chi^{\dagger}] \det[\phi^{\dagger},\chi]\big]^{-1}$ correspond to Nielsen's $(\det \gamma \cdot D)^{-2} \delta(\gamma^{\rho}\psi_{\rho}-b)\delta(\bar \psi_\lambda \gamma^\lambda -\bar b)$, since $\chi$ and $\chi^{\dagger}$ are the
generators of the fermionic gauge transformation.  The FP ghost contribution to the chiral anomaly will then be -1 by his argument, and the calculation
to be done is to find the anomaly contribution from the fermion fields $\psi_{\mu}, \,\lambda, \, \Lambda$, with a gauge fixing term
\begin{equation}\label{gaugefix}
\Delta {\cal L}= -\zeta \bar \psi_{\mu}\gamma^{\mu} \gamma^\nu D_{\nu} \gamma^{\rho} \psi_{\rho}
\end{equation}
added to the action.

The Lagrangian ${\cal L}$ appearing in Eq. \eqref{pathint} can be read off from Eq. \eqref{covaction}.  Using the identity
$\epsilon^{\mu\eta\nu\rho}\gamma_5\gamma_{\eta}=i\gamma^{\mu\nu\rho}$, the Lagrangian takes the form
\begin{align}\label{lagrangian}
{\cal L}=& -[\bar \psi_{\mu}\gamma^{\mu\nu\rho}D_{\nu}\psi_{\rho}+(g/2)(-\bar \Lambda \gamma^{\mu\nu\rho}F_{\mu\nu} \psi_{\rho}
+ \bar \psi_{\mu} \gamma^{\mu\nu\rho}F_{\nu\rho} \Lambda-\bar \Lambda \gamma^{\mu\nu\rho}F_{\nu\rho} D_{\mu} \Lambda)]  \cr
-&\bar \lambda \gamma^\nu D_\nu \lambda +m(\bar \lambda \gamma^\nu \psi_\nu-\bar{\psi}_\nu \gamma^\nu \lambda +\bar \lambda \gamma^\nu D_\nu \Lambda -\bar \Lambda \overleftarrow{D}_\nu\gamma^{\nu} \lambda)~~~.\cr
\end{align}
For the later derivation of Feynman rules, we expand the gauge-fixed  Lagrangian density ${\cal L}+ \Delta {\cal L}$ in powers of the coupling $g$,
\begin{align}\label{powerexp}
{\cal L}+\Delta {\cal L}=& {\cal L}^{(0)}+g{\cal L}^{(1)}+g^2 {\cal L}^{(2)}~~~,\cr
{\cal L}^{(0)}=&  -\bar \psi_{\mu}\gamma^{\mu\nu\rho}\partial_{\nu}\psi_{\rho}   -\bar \lambda \gamma^\nu \partial_\nu \lambda
-\zeta \bar \psi_{\mu}\gamma^{\mu} \gamma^\nu \partial_{\nu} \gamma^{\rho} \psi_{\rho}
+m(\bar \lambda \gamma^\nu \psi_\nu-\bar{\psi}_\nu \gamma^\nu \lambda +\bar \lambda \gamma^\nu \partial_\nu \Lambda -\bar \Lambda \overleftarrow{\partial}_\nu\gamma^{\nu} \lambda)~~~,\cr
{\cal L}^{(1)}=& A_\nu U^\nu + F_{\alpha\beta}V^{[\alpha\beta]}   ~~~,\cr
{\cal L}^{(2)}=&  F_{\nu\rho}A_{\mu} W^{[\nu\rho\mu]}     ~~~,\cr
\end{align}
where we have introduced the definitions
\begin{align}\label{UVWdef}
U^\nu =& -\bar \psi_{\mu}\gamma^{\mu\nu\rho} \psi_{\rho} -\bar \lambda \gamma^\nu \lambda + m(\bar \lambda \gamma^{\nu} \Lambda +
\bar \Lambda \gamma^\nu \lambda) -\zeta \bar \psi_{\mu}\gamma^{\mu} \gamma^\nu  \gamma^{\rho} \psi_{\rho}    ~~~,\cr
V^{[\alpha\beta]}=&\frac{1}{2}(\bar \Lambda \gamma^{\alpha\beta\rho}\psi_\rho-\bar{\psi}_\mu\gamma^{\alpha\beta\mu}\Lambda + \bar \Lambda \gamma^{\alpha\beta\tau}\partial_{\tau}\Lambda)~~~,\cr
W^{[\nu\rho\mu]} =&\frac{1}{2} \bar \Lambda  \gamma^{\mu\nu\rho} \Lambda ~~~.\cr
\end{align}
We see that in addition to simple vector vertices where $A_\nu$ couples to a vector current, there are vertices with $F_{\mu\nu}$ coupling
to a rank two antisymmetric tensor current, and with $F_{\nu\rho}A_\mu$ coupling to a rank three antisymmetric tensor current.

Before proceeding to Feynman rules, let us give the relation between the quantities just defined and the Noether currents.  The Noether vector
current is obtained by making the substitutions
\begin{equation}\label{vectornoether}
\psi_\rho \to \exp(\theta) \psi_{\rho}~,~~\lambda \to  \exp(\theta) \lambda~,~~ \Lambda \to  \exp(\theta) \Lambda~~~,
\end{equation}
with $\theta^{\dagger}=-\theta$, and picking out the coefficient of $\partial_\sigma \theta$.  This gives the Noether vector current ${\cal V}^\sigma$
given by
\begin{align}\label{Vdef}
{\cal V}^\sigma=&{\cal V}^{(0)\sigma}+{\cal V}^{(1)\sigma}~~~,\cr
{\cal V}^{(0)\sigma}=&U^\sigma =-\bar{\psi}_{\mu} \gamma^{\mu\sigma\rho}\psi_\rho  -\zeta \bar{\psi}_\mu \gamma^\mu \gamma^\sigma \gamma^\rho \psi_\rho - \bar \lambda \gamma^\sigma \lambda
+m(\bar \lambda \gamma^{\sigma} \Lambda + \bar \Lambda \gamma^\sigma \lambda)~~~,\cr
{\cal V}^{(1)\sigma}=& gF_{\nu \rho} W^{[\nu \rho\sigma]}=\frac{g}{2} \bar \Lambda \gamma^{\sigma \nu\rho}F_{\nu\rho} \Lambda ~~~.\cr
\end{align}
Similarly, making the substitution of Eq. \eqref{vectornoether} with $\theta$ replaced by $-\gamma_5 \theta$, we find the Noether axial-vector
current ${\cal A}^\sigma$ given by
\begin{align}\label{Adef}
{\cal A}^\sigma=&{\cal A}^{(0)\sigma}+{\cal A}^{(1)\sigma}~~~,\cr
{\cal A}^{(0)\sigma}=&\bar{\psi}_{\mu} \gamma^{\mu\sigma\rho}\gamma_5 \psi_\rho  +\zeta \bar{\psi}_\mu \gamma^\mu \gamma^\sigma \gamma^\rho \gamma_5 \psi_\rho + \bar \lambda \gamma^\sigma\gamma_5 \lambda
-m(\bar \lambda \gamma^{\sigma}\gamma_5 \Lambda + \bar \Lambda \gamma^\sigma \gamma_5 \lambda)~~~,\cr
{\cal A}^{(1)\sigma}=& -\frac{g}{2} \bar \Lambda \gamma^{\sigma \nu\rho}\gamma_5 F_{\nu\rho} \Lambda~~~.\cr
\end{align}
showing that the axial current includes a piece with a direct coupling of the vector field through $F_{\nu\rho}$.
One can check that the Noether currents just defined are self-adjoint, ${\cal V}^\sigma = ({\cal V}^\sigma)^\dagger$ ~,~~
${\cal A}^\sigma = ({\cal A}^\sigma)^\dagger$, and by a lengthy calculation using the Euler-Lagrange equations following from the
action of Eq. \eqref{covaction}, with the gauge fixing action added, one can check that the Noether currents are formally conserved,
$\partial_\sigma  {\cal V}^\sigma =\partial_\sigma  {\cal A}^\sigma =0.$

\section{Feynman rules for propagators and vertices }
\label{section_Feynman_rules}

Let us next derive the Feynman rules.  We introduce Fourier transforms of the fields
\begin{align}\label{fourier}
\psi_{\mu}(x)=&\frac{1}{(2\pi)^4} \int d^4k e^{ik\cdot x}\psi_{\mu}[k]~~~,\cr
\lambda(x)=&\frac{1}{(2\pi)^4} \int d^4k e^{ik\cdot x}\lambda[k]~~~,\cr
\Lambda(x)=&\frac{1}{(2\pi)^4} \int d^4k e^{ik\cdot x}\Lambda[k]~~~,\cr
S=&\frac{1}{(2\pi)^4} \int d^4kS[k]~~~,\cr
\end{align}
and expand
\begin{equation}\label{expandS}
S[k]=S^{(0)}[k]+g S^{(1)}[k]+g^2 S^{(2)}[k]
\end{equation}
corresponding to the Lagrangian density expansion of Eq.  \eqref{powerexp}.
Then for the kinetic term $S^{(0)}[k]$ we find
\begin{equation}\label{Skin}
S^{(0)}[k]=\Big( \bar{\psi}_{\mu}[k] \, \bar{\lambda}[k] \, \bar\Lambda[k] \Big) {\cal M}
\left(
  \begin{array}{c}
    \psi_{\rho}[k] \\
    \lambda[k] \\
    \Lambda[k] \\
  \end{array}
\right) \;~~~.
\end{equation}
For the matrix ${\cal M}$ we have
\begin{equation}\label{matrixM}
{\cal M}=   \left[ \begin{array} {c c c }
-i[(\frac{1}{2}+\zeta)\gamma^\mu\slashed{k}\gamma^\rho-\frac{1}{2}\gamma^\rho\slashed{k}\gamma^{\mu}]  &~~~-m\gamma^\mu &~~~ 0   \\
m\gamma^{\rho}  & ~~~-i\slashed{k} & ~~~ im\slashed{k}  \\
 0 & ~~~im\slashed{k}  &~~~  0  \\
 \end{array}\right]~~~.
\end{equation}
Defining the propagator ${\cal N}$ as the inverse of ${\cal M}$,
\begin{equation}\label{inverse1}
{\cal M} {\cal N}= \left[ \begin{array} {c c c }
\delta^{\mu}_{\sigma}  & 0 & 0   \\
0 & 1 & 0  \\
 0 & 0  &  1  \\
 \end{array}\right]~~~,
\end{equation}
and writing
\begin{equation}\label{Nform}
{\cal N}=\left[ \begin{array} {c c c }
N_{1\rho\sigma} & N_{2\rho} & N_{3\rho}  \\
N_{4\sigma}  & N_5 & N_6  \\
N_{7\sigma} &N_8  & N_9 \\
 \end{array}\right]~~~,
\end{equation}
we find the following solution for the matrix elements of ${\cal N}$,
\begin{align}\label{Nmatrixelts}
N_{1\rho\sigma}=&-\frac{i}{2k^2}[\gamma_\sigma\slashed{k}\gamma_\rho-\frac{1}{k^2}(4+\frac{2}{\zeta})k_\rho k_\sigma \slashed{k} ]~~~, \cr
N_{2\rho}=&0~~~,\cr
N_{3\rho}=&\slashed{k}\frac{1}{\zeta (k^2)^2} k_{\rho}~~~,\cr
N_{4\sigma}=&0~~~,\cr
N_5=&0~~~,\cr
N_6=&\slashed{k}\frac{1}{i m k^2}~~~,\cr
N_{7\sigma}=&\slashed{k}\frac{-1}{\zeta (k^2)^2} k_{\sigma}~~~,\cr
N_8=&\slashed{k}\frac{1}{i m k^2}~~~,\cr
N_9=&\frac{\slashed{k}}{i k^2}\left(\frac{1}{m^2} -\frac{1}{\zeta k^2}\right) ~~~.\cr
\end{align}

The terms  $S^{(1)}[k]$ and $S^{(2)}[k]$  in  Eq. \eqref{expandS} give the vertex Feynman rules.
The vertices corresponding to ${\cal V}^{(0)\sigma}$ and ${\cal A}^{(0)\sigma}$ are
\begin{align}\label{vertices}
 {\cal V}^{(0)\sigma}=&\left[ \begin{array} {c c c }
-[(\frac{1}{2}+\zeta)\gamma^\mu\gamma^\sigma\gamma^\rho-\frac{1}{2}\gamma^\rho\gamma^\sigma\gamma^{\mu}]  &~~~0 &~~~ 0   \\
0  & ~~~-\gamma^\sigma & ~~~ m\gamma^\sigma  \\
 0 & ~~~m\gamma^\sigma  &~~~  0  \\
 \end{array}\right]~~~,\cr
{\cal A}^{(0)\sigma}=&\left[ \begin{array} {c c c }
[(\frac{1}{2}+\zeta)\gamma^\mu\gamma^\sigma\gamma^\rho-\frac{1}{2}\gamma^\rho\gamma^\sigma\gamma^{\mu}]\gamma_5  &~~~0 &~~~ 0   \\
0  & ~~~\gamma^\sigma\gamma_5  & ~~~ -m\gamma^\sigma\gamma_5   \\
 0 & ~~~-m\gamma^\sigma\gamma_5   &~~~  0  \\
 \end{array}\right]~~~,\cr
\end{align}
and obey the Ward identitites
\begin{align}\label{ward}
ik_\sigma {\cal V}^{(0)\sigma}=&{\cal M}(k+p)-{\cal M}(p)={\cal N}^{-1}(k+p)-{\cal N}^{-1}(p)~~~,\cr
-ik_\sigma {\cal A}^{(0)\sigma}=&{\cal M}(k+p)\gamma_5+\gamma_5{\cal M}(p)={\cal N}^{-1}(k+p)\gamma_5+\gamma_5{\cal N}^{-1}(p)~~~.\cr
\end{align}

The corresponding Feynman rules for $V^{[\alpha\beta]}$ and $W^{[\nu\rho\mu]}$ are
\begin{align}\label{VandW}
V^{[\alpha\beta]}=&\left[ \begin{array} {c c c }
0 &~~~0 &~~~-\frac{1}{2}\gamma^{\mu\alpha\beta}  \\
0  &0 & 0 \\
 \frac{1}{2}\gamma^{\alpha\beta\rho}& 0 & \frac{i}{2}\gamma^{\alpha\beta\tau}k_{\tau} \\
 \end{array}\right]~~~,\cr
W^{[\nu\rho\mu]}=&\left[ \begin{array} {c c c }
0  &~~~0 &~~~ 0   \\
0  &0 & 0 \\
 0 & 0  &~~~  \frac{1}{2}\gamma^{\nu\rho\mu}  \\
 \end{array}\right]~~~,\cr
\end{align}
where $k$ is the four-momentum of the $\Lambda$ entering or leaving a vertex where $F_{\alpha\beta}V^{[\alpha\beta]}$ couples.

\section{Enumeration of diagrams}
\label{section_enumeration_diagrams}

\begin{figure}%[tbp]%\begin{flushleft}
\begin{center}
\includegraphics[width=.8\textwidth]{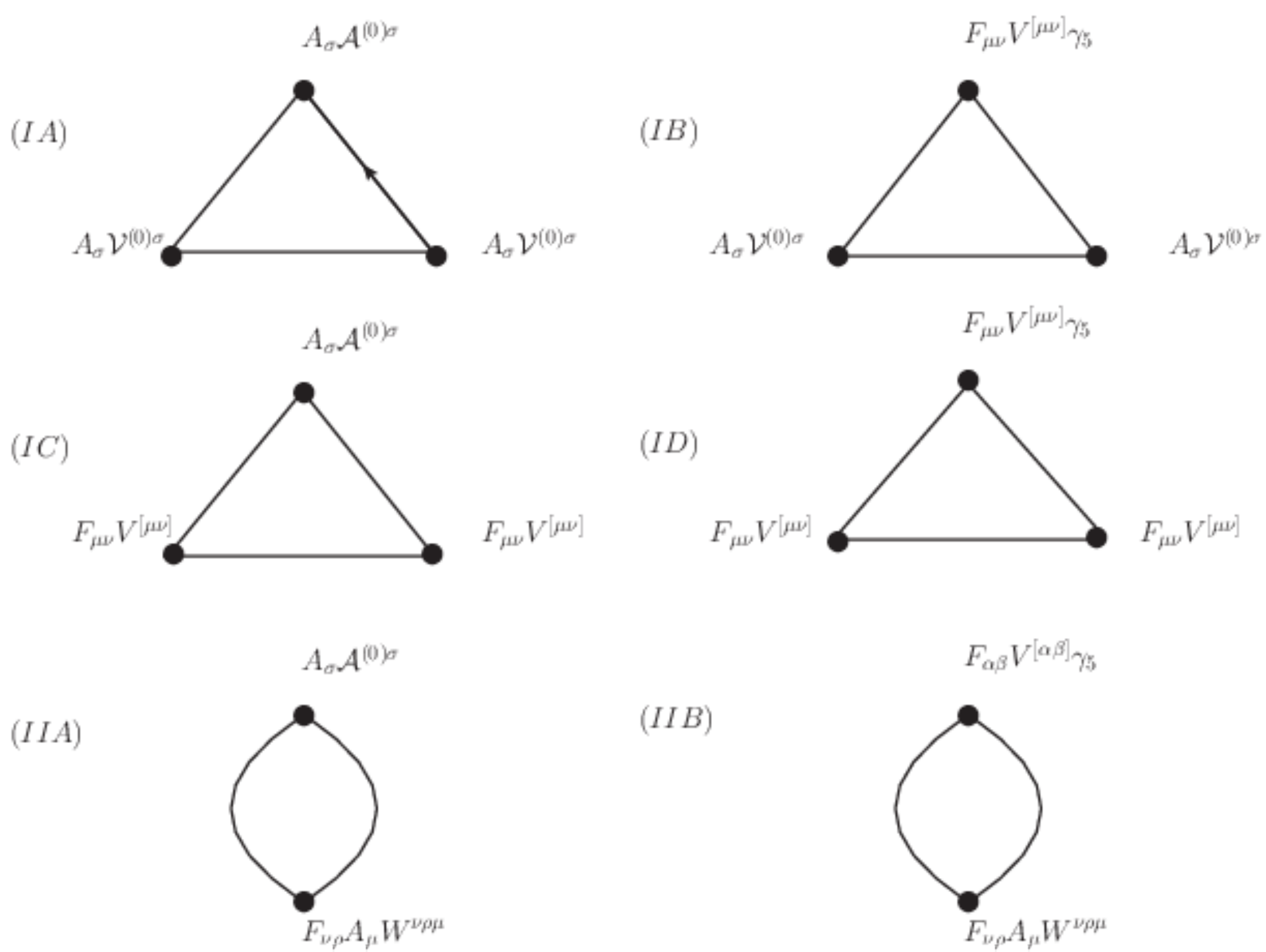}
\end{center}
\caption{Summary of contributing diagrams.}
%\end{flushleft}
\label{fig_diag_cont}
\end{figure}

The issue of anomaly cancellation arises when a left (or right) chiral current is gauged, since when anomalies in the gauge gluon three point function are
not cancelled there are non-renormalizable infinities \cite{gross}, \cite{bouchiat}. The general three-point function for chiral currents gets contributions
from diagrams with one axial-vector coupling and two vector couplings, and diagrams with three axial-vector couplings (diagrams with two axial-vector couplings and one
vector coupling, as well as diagrams with three vector couplings, vanish by charge conjugation symmetry). Since the anomaly associated with three axial-vector couplings
is known on symmetry grounds to be $\frac{1}{3}$ that of the anomaly associated with one axial-vector and two vector couplings, it suffices to compute the latter in a
vector-like theory, and then to supply the appropriate symmetry factors to get the anomaly in a chiral theory.  Thus, the relevant diagrams for our calculation
are all of those of order $g^3$ with one axial-vector vertex and two vector vertices, with the axial current ${\cal A}^{\sigma}$ of \eqref{Adef} multiplied by
a factor of the coupling $g$, since in a chiral theory this is gauged as well as the vector current.  This leads to the following enumeration of diagrams, as shown
in Fig. \ref{fig_diag_cont}, where we have included the gauge field factors coupling to each vertex.

Diagrams which are labeled type I are triangles.  Diagram IA has a leading order Noether axial-vector current $A_{\sigma}{\cal A}^{(0)\sigma}$ at one vertex and leading order Noether vector currents
$A_{\sigma}{\cal V}^{(0)\sigma}$ at the other two vertices.  Using the freedom to anticommute the factor $\gamma_5$ around in diagrams with massless propagators, the other triangle
diagrams are IB, in which there is one axial-vector vertex $F_{\mu\nu}V^{[\mu\nu]}\gamma_5$ and two vector vertices $A_{\sigma}{\cal V}^{(0)\sigma}$, Diagram IC in which
there is one axial-vector vertex $A_{\sigma}{\cal A}^{(0)\sigma}$ and two vector vertices  $F_{\mu\nu}V^{[\mu\nu]}$, and Diagram ID, in which there is one axial-vector vertex
$F_{\mu\nu}V^{[\mu\nu]}\gamma_5$  and two vector vertices $F_{\mu\nu}V^{[\mu\nu]}$.  In addition, there are diagrams that are two-point functions
at which two gluons couple to one of the vertices, which we label as type II.  Diagram IIA has one axial-vector vertex $A_{\sigma}{\cal A}^{(0)\sigma}$ and one double vector vertex $F_{\nu\rho}A_{\mu}W^{[\nu\rho\mu]}$, and diagram IIB has one axial-vector vertex  $F_{\alpha\beta}V^{[\alpha\beta]}\gamma_5$ and one double vector vertex $F_{\nu\rho}A_{\mu}W^{[\nu\rho\mu]}$.  All other possibilities can be reduced to the ones just enumerated by moving the factor $\gamma_5$ around inside the fermion loop
trace.

\section{Anomaly arising from the leading order Noether current triangle}
\label{section_anomaly_triangle}

In this section we evaluate the anomaly arising from diagram IA.  We first review the calculation of the standard anomaly for spin-$\frac{1}{2}$ following
the treatment in \cite{adler2}, and then do the analogous calculation for the diagram with leading order Noether currents at the three vertices.

\subsection{The standard spin-$\frac{1}{2}$ chiral anomaly by the shift method}

\begin{figure}%[tbp]
%\begin{flushleft}
\begin{center}
\includegraphics[width=.8\textwidth]{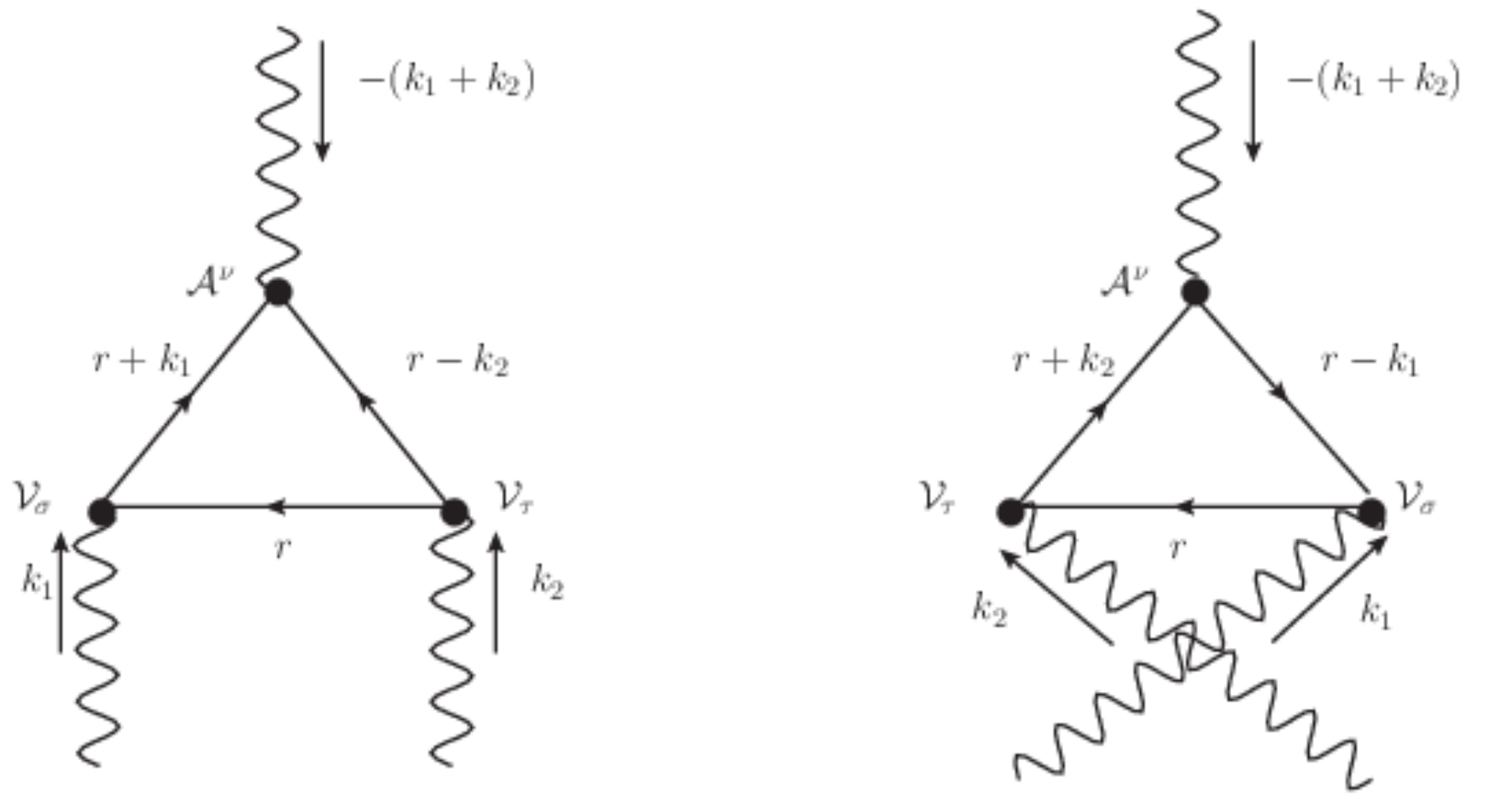}
\end{center}
\caption{Generic triangle diagrams.}
%\end{flushleft}
\label{fig_diagrams_IA}
\end{figure}

The generic triangle diagrams with one axial-vector vertex and two vector vertices are shown in Fig. \ref{fig_diagrams_IA}.

 For the case of the standard spin-$\frac{1}{2}$  anomaly,
the vertex ${\cal A}$ with incoming momentum $-(k_1+k_2)$ is $\gamma^{\nu}\gamma_5$, and the vector vertices ${\cal V}$ with incoming momenta $k_1$ and $k_2$ are $-i\gamma_{\sigma}$ and $-i\gamma_{\tau}$ respectively.  For the corresponding amplitude, we find
\begin{align}\label{amplitude1}
{\cal T}_{\sigma\tau}^{\nu}=& \int \frac{d^4 r}{(2\pi)^4} (-1) \rm{tr} \left[\frac{i}{\slashed{r}+\slashed{k_1}}(-i\gamma_{\sigma})\frac{i}{\slashed{r}}(-i\gamma_{\tau})
\frac{i}{\slashed{r}-\slashed{k_2}}\gamma^{\nu}\gamma_5\right] \cr
+& \int \frac{d^4 r}{(2\pi)^4} (-1) \rm{tr} \left[\frac{i}{\slashed{r}+\slashed{k_2}}(-i\gamma_{\tau})\frac{i}{\slashed{r}}(-i\gamma_{\sigma})
\frac{i}{\slashed{r}-\slashed{k_1}}\gamma^{\nu}\gamma_5\right]~~~. \cr
\end{align}
Forming the axial-vector divergence $-(k_1+k_2)_{\nu}{\cal T}_{\sigma\tau}^{\nu}$, and substituting $-(\slashed{k_1}+\slashed{k_2})\gamma_5 = (\slashed{r}-\slashed{k_2})\gamma_5 + \gamma_5
(\slashed{r}+\slashed{k_1})$  into the first line and $-(\slashed{k_1}+\slashed{k_2})\gamma_5 = (\slashed{r}-\slashed{k_1})\gamma_5 + \gamma_5
(\slashed{r}+\slashed{k_2})$ into the second line, one gets a sum of four terms, each of which contains only $k_1$ or $k_2$ but not both, and hence vanishes,
since there are not enough external momentum factors to form the pseudoscalar $\epsilon^{\tau\sigma\mu\nu} k_{1\, \mu} k_{2\, \nu}$.  Hence with the the chosen routing
of momenta in the triangle, the axial-vector divergence vanishes.  Since the sum of the two diagrams is symmetric under interchange of the vector vertices, it suffices
to test the single vector divergence $k_1^{\sigma} {\cal T}_{\sigma\tau}^{\nu}$, by substituting $\slashed{k_1}=(\slashed{r}+\slashed{k_1}) - \slashed{r}$ into the
first line and $\slashed{k_1}= \slashed{r}- (\slashed{r}-\slashed{k_1})$  into the second line.  This gives a sum of four terms, two of which contain only $k_2$, and hence
vanish, leaving the other two terms,
\begin{equation}\label{shift1}
k_1^{\sigma} {\cal T}_{\sigma\tau}^{\nu}=i\int \frac{d^4 r}{(2\pi)^4}{\rm tr} \left[ \frac{1}{(\slashed{r}+\slashed{k_1})} \gamma_{\tau} \frac{1}{{\slashed{r}-\slashed{k_2}}} \gamma^{\nu} \gamma_5 - \frac{1}{(\slashed{r}+\slashed{k_2})} \gamma_{\tau} \frac{1}{{\slashed{r}-\slashed{k_1}}} \gamma^{\nu} \gamma_5\right]~~~.
\end{equation}
If we could make the shift of integration variable $r \to r+k_2-k_1$ in the first term of Eq. \eqref{shift1}, the two terms would cancel, but this shift is not permitted
inside a linearly divergent integral.  Following Jackiw \cite{jackiwlectures} we proceed as follows.  Taking $k_1-k_2$ to be infinitesimal, and rationalizing Feynman
denominators, we can write
Eq. \eqref{shift1} as
\begin{align}\label{shift2}
k_1^{\sigma} {\cal T}_{\sigma\tau}^{\nu}
\simeq  & i (k_1-k_2)_{\kappa}\int \frac{d^4 r}{(2\pi)^4} \frac{\partial}{\partial r_{\kappa}} \left[\frac{{\rm tr}\big((\slashed{r}+\slashed{k_2}) \gamma_{\tau} (\slashed{r}-\slashed{k_1}) \gamma^{\nu} \gamma_5\big)}{(r+k_2)^2(r-k_1)^2}\right] ~~~.\cr
\end{align}
Let us now make the usual Wick rotation to a Euclidean integration region for $r$, which introduces an overall factor of $i$, and use Stokes theorem, which
for a Euclidean four-dimensional integration over a volume $V$ bounded by a surface $S$ states that
\begin{equation}\label{stokes1}
\int_{\rm V}  d^4r \frac{\partial}{ \partial r_{\kappa}} f(r) = \int_{S} dS^{\kappa} f(r)~~~.
\end{equation}
Applying Eq. \eqref{stokes1} to Eq. \eqref{shift2}, we have
\begin{align}\label{shift3}
k_1^{\sigma} {\cal T}_{\sigma\tau}^{\nu}
\simeq  &\frac{ -1}{(2\pi)^4} (k_1-k_2)_{\kappa}\int_S dS^{\kappa}\frac{{\rm tr} \left[ (\slashed{r}+\slashed{k_2}) \gamma_{\tau} (\slashed{r}-\slashed{k_1}) \gamma^{\nu} \gamma_5\right]}{(r+k_2)^2(r-k_1)^2}    ~~~.\cr
\end{align}
The trace in the numerator can be simplified to ${\rm tr}[\big((\slashed{k_1}+\slashed{k_2})\gamma_{\tau} \slashed{r}-\slashed{k_2} \gamma_{\tau} \slashed{k_1}\big)
\gamma^{\nu}\gamma_5]$.  Taking now the surface $S$ to be a large three-sphere of radius $R$, the denominator $(r+k_2)^2(r-k_1)^2\simeq R^4$ and so can be pulled
outside the integral. Since the volume of the sphere is $2\pi^2 R^3$, and noting that $dS^{\kappa}$ is a vector parallel
to $r^{\kappa}$, the $r$-independent term in the numerator averages to zero, while  $\slashed{r}$ averages to $R(\gamma^{\kappa}/4)$, giving
\begin{equation}\label{average}
\int_S dS^{\kappa}  {\rm tr}[\big((\slashed{k_1}+\slashed{k_2})\gamma_{\tau} \slashed{r}-\slashed{k_2} \gamma_{\tau} \slashed{k_1}\big)
\gamma^{\nu}\gamma_5]=2\pi^2 R^4  {\rm tr}[(\slashed{k_1}+\slashed{k_2})\gamma_{\tau} (\gamma^{\kappa}/4) \gamma^{\nu}\gamma_5]~~~.
\end{equation}
Thus the $R$ factors cancel out as the sphere radius approaches $\infty$, and we find for the
vector vertex anomaly
\begin{align}\label{finalanomaly}
k_1^{\sigma} {\cal T}_{\sigma\tau}^{\nu}=&\frac{ -g^2}{(2\pi)^4} (k_1-k_2)_{\kappa}2\pi^2{\rm tr}[(\slashed{k_1}+\slashed{k_2})\gamma_{\tau} (\gamma^{\kappa}/4) \gamma^{\nu}\gamma_5]\cr
= &\frac{g^2}{16 \pi^2} {\rm tr}[\slashed{k_1}\gamma_{\tau} \slashed{k_2} \gamma^{\nu}\gamma_5]~~~.\cr
\end{align}
When vector vertex conservation is enforced by adding a polynomial to the amplitude, Eq. \eqref{finalanomaly} yields the usual answer for the axial-vector anomaly.
In comparing with the coupled model calculation that follows, it suffices to use the expression in Eq. \eqref{shift1} for the standard spin-$\frac{1}{2}$ anomaly, so we will not repeat the steps of Eqs. \eqref{stokes1} through \eqref{finalanomaly}.

\subsection{The anomaly arising from diagram IA}

Referring to Fig. \ref{fig_diagrams_IA}, for diagram IA
the axial-vector vertex ${\cal A}$ with incoming momentum $-(k_1+k_2)$ is ${\cal A}^{(0)}$, and the vector vertices ${\cal V}$ with incoming momenta $k_1$ and $k_2$ are
$i{\cal V}^{(0)}_\sigma$ and $i{\cal V}^{(0)}_\tau$ respectively.  For the corresponding amplitude, we find
\begin{align}\label{newamplitude}
\tilde{{\cal T}}_{\sigma\tau}^{\nu}=& \int \frac{d^4 r}{(2\pi)^4}  \rm{tr} [{\cal N}(r+k_1) {\cal V}^{(0)}_{\sigma} {\cal N}(r) {\cal V}^{(0)}_{\tau} {\cal N}(r-k_2){\cal A}^{(0)\nu} \cr  +&{\cal N}(r+k_2) {\cal V}^{(0)}_{\tau} {\cal N}(r) {\cal V}^{(0)}_{\sigma}{\cal N}(r-k_1){\cal A}^{(0)\nu}] ~~~.\cr
\end{align}
 Contracting with $i(k_1+k_2)_{\nu}$ to test the axial divergence, and using the respective  Ward identities
\begin{align}\label{newward1}
i(k_1+k_2)_\nu {\cal A}^{(0)\nu}=&{\cal N}^{-1}(r-k_2)\gamma_5+\gamma_5{\cal N}^{-1}(r+k_1)\cr~~~
=&{\cal N}^{-1}(r-k_1)\gamma_5+\gamma_5{\cal N}^{-1}(r+k_2)~~~,\cr
\end{align}
in the first and second lines of Eq. \eqref{newamplitude}, we get again a sum of four terms, each of which contains only $k_1$ or $k_2$ and so vanish.  So the
axial-vector divergence vanishes.  Contracting with ${k_1}^{\sigma}$ to test the vector divergence, and using the respective Ward identities
\begin{align}\label{newward2}
ik_1^{\sigma} {\cal V}^{(0)}_{\sigma}=&{\cal N}^{-1}(r+k_1)-{\cal N}^{-1}(r)~~~,\cr
=&{\cal N}^{-1}(r)-{\cal N}^{-1}(r-k_1)~~~,\cr
\end{align}
in the first and second lines of Eq. \eqref{newamplitude}, we get a sum of four terms, two of which contain only $k_2$ and vanish, leaving the other two terms
\begin{equation}\label{newshift1}
k_1^{\sigma}\tilde{{\cal T}}_{\sigma\tau}^{\nu}=i\int \frac{d^4 r}{(2\pi)^4}\rm{tr} [{\cal N}(r+k_1)  {\cal V}_{\tau} {\cal N}(r-k_2){\cal A}^{\nu}
-{\cal N}(r+k_2) {\cal V}_{\tau} {\cal N}(r-k_1){\cal A}^{\nu}] ~~~.
\end{equation}
Again, we see that if we could make a shift of integration variable $r \to r+k_2-k_1$ in the first term of Eq. \eqref{newshift1}, the two terms would cancel, but as before this shift is not permitted inside a linearly divergent integral.  To proceed further we focus on the first term in Eq. \eqref{newshift1},
substitute the propagator and vertex matrices from Eqs. \eqref{Nform} and \eqref{vertices}, multiply out, and take the overall trace.
Writing Eq. \eqref{newshift1} as
\begin{align}\label{newshift2}
k_1^{\sigma}\tilde{{\cal T}}_{\sigma\tau}^{\nu}=&i\int \frac{d^4 r}{(2\pi)^4} \tilde S~~~,\cr
\tilde S\equiv&\rm{tr} [{\cal N}(r+k_1)  {\cal V}_{\tau} {\cal N}(r-k_2){\cal A}^{\nu}-(k_1 \leftrightarrow k_2)]   ~~~,\cr
\end{align}
and abbreviating
$s\equiv r+k_1$, $d\equiv r-k_2$, $(V^{\sigma})^{\mu\rho}\equiv (\frac{1}{2}+\zeta)\gamma^\mu\gamma^\sigma\gamma^\rho-\frac{1}{2}\gamma^\rho\gamma^\sigma\gamma^{\mu} $,
we find for the explicitly shown term in $\tilde S$ the expression
\begin{align}\label{s}
\tilde S[{\rm explicitly ~shown ~term}]=&{\rm tr}[-N_{1\rho\sigma}(s)V_\tau^{\sigma\alpha}N_{1\alpha\beta}(d)(V^{\nu})^{\beta \rho} \gamma_5\cr
 -& m^2 N_6(s)\gamma_\tau N_6(d)\gamma^\nu \gamma_5-m^2 N_8(s)\gamma_\tau N_8(d) \gamma^\nu \gamma_5] ~~~.\cr
\end{align}
Substituting $N_6$ and $N_8$ from Eq. \eqref{Nmatrixelts} we see that the factors of $m$ cancel, leaving as the sum of the second and third terms in Eq. \eqref{s}
\begin{equation}\label{sum}
\frac{2}{s^2d^2}{\rm tr}[\slashed{s}\gamma_\tau \slashed{d}\gamma^\nu \gamma_5]~~~.
\end{equation}
When substituted into Eq. \eqref{newshift2} this gives exactly twice the first term in Eq. \eqref{shift1}, corresponding to a factor of 2 times the standard spin-$\frac{1}{2}$ anomaly.

The first term in Eq. \eqref{s} is more complicated in structure.  We have evaluated it two different ways.  By using the cyclic invariance of the trace, this term
can be evaluated algebraically for general gauge parameter $\zeta$ using the identities in Appendix \ref{appendix_A}, with the result
\begin{equation}\label{first}
\frac{4}{s^2d^2}\Big[1+\frac{1}{16}(\frac{1}{2}+\zeta)\Sigma\Big] {\rm tr}[\slashed{s}\gamma_\tau \slashed{d}\gamma^\nu \gamma_5]~~~,
\end{equation}
with $\Sigma$ given by
\begin{equation}\label{sigma}
\Sigma=-16-16(\frac{1}{2}+\zeta)+\frac{4}{\zeta}[-2+8(\frac{1}{2}+\zeta)+8(\frac{1}{2}
+\zeta)^2]+\frac{4}{\zeta^2}[3(\frac{1}{2}+\zeta)-4(\frac{1}{2}+\zeta)^2-4(\frac{1}{2}+\zeta)^3]\equiv 0~~~.
\end{equation}
Thus, the anomaly from diagram IA is independent of the gauge fixing parameter $\zeta$, and adding Eq. \eqref{first} to Eq. \eqref{sum} and substituting the
total into Eq. \eqref{newshift2} gives six times the first term in Eq. \eqref{shift1}.  So the diagram IA contribution to the chiral anomaly is a factor of 6
times  the standard spin-$\frac{1}{2}$ anomaly.  As a check on this calculation, we also evaluated the first term in Eq. \eqref{s} in the gauge $\zeta=-\frac{1}{2}$, which
eliminates many terms from the calculation, and used the FEYNCALC package of Mathematica \cite{feyncalc}
 to evaluate the Dirac matrix trace, with the same result of six times the standard anomaly.

\section{Anomaly contributions arising from the remaining diagrams}
\label{section_remaining_diagrams}

We turn next to calculating the anomaly contributions coming from the remaining diagrams in Fig. \ref{fig_diag_cont}.  This is facilitated by the observation that since the anomaly is topological in nature, it cannot depend on continuously variable parameters such as the coupling mass $m$ and the
gauge-fixing parameter $\zeta$.  We have seen an example of this in the preceding section, where the apparent $m$ and $\zeta$ dependence
cancelled away in the calculation of the diagram IA anomaly.  In the calculations of this section, after multiplying all vertex and
propagator factors, we shall take the limit as $m$ and $\zeta$  become infinite, dropping terms which vanish in this limit, and keeping
only terms which remain finite (we  find no growing terms).  This leaves only a few remaining pieces to evaluate algebraically to get
the anomaly contribution.

The result of this calculation is that all of the remaining diagrams contribute zero to the chiral anomaly.  We enumerate them one by one,
giving for each the reason why they give a null contribution.

\begin{itemize}
\item   Diagram ID.  Since all vertices have a field strength factor $F_{\alpha \beta}[k]= i(k_\alpha A_\beta-k_\beta A_\alpha)$, which vanishes
when $A_\alpha = k_\alpha$, this diagram is conserved at all three vertices.  Note also that it is of order $k_1^2k_2$  and $k_2^2k_1$ in
external momenta, so is a higher order polynomial than the anomaly, which is of order $k_1k_2$.
\item   Diagram IC.  For the same reason, it is conserved at the two vector vertices containing $F$ factors.  Taking the divergence at
the axial-vector vertex and using the Ward identities gives a difference of terms which differ by a shift of $k_1$ or $k_2$, plus extra pieces of
order $k_1^2 k_2$ or $k_2^2 k_1$.  Because the shifted terms still each contain a factor $k_1k_2$, the result of the shift is of order $k_1^2 k_2$ or $k_2^2 k_1$, so cannot give an anomaly.  Another reason for a null result is that the terms which are shifted all vanish as $m$ and
$\zeta$ become infinite; there is no contribution that remains nonzero.
\item Diagram IB.  This is conserved at the vertex containing $F$.  Taking a divergence at either of the other two vertices
and substituting the Ward identity gives a difference of terms that differ by a shift of $k_2$ or $k_2-k_1$,  plus a remainder of
order $k_1k_2$ that vanishes as $m$ and $\zeta$ approach infinity.  The shift terms, after dropping terms that are cubic or higher order in $k_{1,2}$, or that vanish as $m$ and $\zeta$ approach infinity, contains a finite part which inside the trace has a factor of either $\gamma^\rho\big(\gamma_\sigma \slashed{r} \gamma_\rho-(4/r^2)r_\sigma r_\rho \slashed{r}\big) =0$ or  $\big(\gamma_\sigma \slashed{r} \gamma_\rho-(4/r^2)r_\sigma r_\rho \slashed{r}\big)\gamma^\sigma =0$. (This is the identity of Eq. (72) of \cite{adler2}).  Hence the anomaly from diagram IB is zero.
\item Diagrams IIA and IIB.  These diagrams approach zero as as $m$ and $\zeta$ approach infinity, with no nonzero remainder, so contribute
zero to the anomaly.
\end{itemize}

Our conclusion is that the remaining diagrams IB, IC, ID and IIA, IIB all contribute zero to the anomaly.  So the anomaly is given entirely by
diagram IA, which gives 6 times the standard spin-$\frac{1}{2}$ anomaly, together with the ghost contribution of $-1$ times the standard
anomaly, giving a total of 5 times the standard anomaly.  As noted in the Introduction, this agrees with the result obtained in \cite{adler2}
when the second class constraint determinant is exponentiated by introducing a non-propagating ghost.

\section{Discussion}
\label{section_Discussion}

The result for the coupled model anomaly of $5$ times the standard anomaly differs from what one would get by naive counting in the uncoupled
model.  In the model with $m=0$ (which as noted has singularities that prevent a perturbative gauging), one would naively count an anomaly of
1 for the spin-$\frac{1}{2}$ field and an anomaly of 3 for the spin-$\frac{3}{2}$ field \cite{naive}, giving a total anomaly of $4$. We see that this naive anomaly counting result cannot be carried over to the coupled model.  This implies that the anomaly counting argument in the non-Abelian $SU(8)$ gauge model of \cite{adler3}, which is the progenitor of the Abelianized coupled model analyzed in \cite{adler2} and here, has to be reexamined.

\section*{ACKNOWLEDGEMENTS}  SLA and PP  wish to thank Marc Henneaux for a useful discussion of ghosts for second class constraints, in which he
favored the non-propagating ghost alternative, as supported now by this paper.  PP acknowledges the warm hospitality of the Centro de Estudios Cient\'ificos (CECs) in Valdivia, Chile, at the final stage of this work.

\appendix

\section{Dirac matrices and identities}
\label{appendix_A}

We follow the conventions used in \cite{adler1}, which agree with those used in the text of Freedman and Van Proeyen \cite{freed}.
Using the flat Minkowskian metric $\eta_{\mu\nu}$, which is $(-,+,+,+)$, the Dirac matrices $\{\gamma_{\mu}\}$ fulfill the Clifford algebra
\begin{equation}\label{Clifford}
\left\{\gamma_{\mu},\gamma_{\nu}\right\}=2\eta_{\mu\nu}.
\end{equation}
In the representation chosen here, they can be written as $4\times4$ matrices in terms of the  Pauli sigma matrices and the $2\times2$ identity matrix $1$,
\begin{eqnarray}\label{gamma_matrices}
\gamma_{0}&=&-\gamma^{0}=\left(
                           \begin{array}{cc}
                             0 & -1 \\
                             1 & 0 \\
                           \end{array}
                         \right) \nonumber \\
\gamma_{i}&=&\gamma^{i}=\left(
                           \begin{array}{cc}
                             0 & \sigma_{i} \\
                             \sigma_{i} & 0 \\
                           \end{array}
                         \right),
\end{eqnarray}
with the $\gamma_{5}$ matrix defined as
\begin{equation}\label{gamma5}
\gamma_{5}=i\gamma_{0}\gamma_{1}\gamma_{2}\gamma_{3}=\left(
                                                       \begin{array}{cc}
                                                         1 & 0 \\
                                                         0 & -1 \\
                                                       \end{array}
                                                     \right) .
\end{equation}

We take the following convention for the Levi-Civita skew-symmetric tensor
\begin{equation}\label{levi-civita}
\epsilon_{0123}=-\epsilon^{0123}=1,
\end{equation}
and, for the spatial Levi-Civita, the identification $\epsilon_{0ijk}=\epsilon_{ijk}$.

Some useful trace properties which can be derived from Eqs. \eqref{Clifford} -- \eqref{gamma5} are
\begin{eqnarray}\label{trace_properties}
\mbox{Tr}(1)&=&4  \; , \nonumber \\
\mbox{Tr}(\mbox{any odd number of $\gamma$'s})&=&0 \;, \nonumber \\
\mbox{Tr}(\gamma_{5})&=&0 \;, \nonumber \\
\mbox{Tr}(\gamma^{\mu}\gamma^{\nu})&=&4\eta^{\mu\nu} \;,   \\
\mbox{Tr}(\gamma^{\mu}\gamma^{\nu}\gamma_{5})&=&0 \;,  \nonumber \\
\mbox{Tr}(\gamma^{\mu}\gamma^{\nu}\gamma^{\rho}\gamma^{\sigma})&=&4\left(\eta^{\mu\nu}\eta^{\rho\sigma}-\eta^{\mu\rho}\eta^{\nu\sigma}+\eta^{\mu\sigma}\eta^{\nu\rho}\right) \; \nonumber,\\
\mbox{Tr}(\gamma^{\mu}\gamma^{\nu}\gamma^{\rho}\gamma^{\sigma}\gamma_{5})&=&-4i\epsilon^{\mu\nu\rho\sigma} \;. \nonumber
\end{eqnarray}

We have also the following contraction formulas,
\begin{eqnarray}\label{contraction_properties}
\gamma^{\mu}\gamma_{\mu}&=&4\;, \nonumber \\
\gamma^{\mu}\gamma^{\nu}\gamma_{\mu}&=&-2\gamma^{\nu} \;, \nonumber \\
\gamma^{\mu}\gamma^{\nu}\gamma^{\rho}\gamma_{\mu}&=&4\eta^{\nu\rho} \;, \\
\gamma^{\mu}\gamma^{\nu}\gamma^{\rho}\gamma^{\sigma}\gamma_{\mu}&=&-2\gamma^{\sigma}\gamma^{\rho}\gamma^{\nu} \;, \nonumber \\
\gamma^{\mu}\gamma^{\nu}\gamma^{\rho}\gamma^{\sigma}\gamma^{\tau}\gamma_{\mu}&=&2\gamma^{\sigma}\gamma^{\rho}\gamma^{\nu}\gamma^{\tau}+2\gamma^{\tau} \gamma^{\nu}\gamma^{\rho}\gamma^{\sigma} \;. \nonumber
\end{eqnarray}

\end{document}